\UseRawInputEncoding
\documentclass[%
 reprint,
 superscriptaddress,
 amsmath, amssymb,
 aps,
 prab,
 longbibliography
]{revtex4-2}

\usepackage{graphicx}
\usepackage{dcolumn}
\usepackage{bm}
\usepackage{multirow}
\usepackage{diagbox}
\usepackage[english]{babel} 
\usepackage{hyperref}
\hypersetup{
    colorlinks=true,
    citecolor=blue,
    filecolor=blue,      
    urlcolor=blue,
}
\usepackage[mathlines]{lineno}

\begin{document}

\preprint{APS/123-QED}

\title{Optical ionization effects in kHz laser wakefield acceleration with few-cycle pulses}

\author{J. Monzac}
\email{josephine.monzac@ensta-paris.fr}
\affiliation{ LOA, CNRS, \'Ecole Polytechnique, ENSTA Paris, Institut Polytechnique de Paris, Palaiseau, France}

\author{S. Smartsev}
\affiliation{ LOA, CNRS, \'Ecole Polytechnique, ENSTA Paris, Institut Polytechnique de Paris, Palaiseau, France}

\author{J. Huijts}%
 \affiliation{ LOA, CNRS, \'Ecole Polytechnique, ENSTA Paris, Institut Polytechnique de Paris, Palaiseau, France}
 
\author{L. Rovige}
\affiliation{ LOA, CNRS, \'Ecole Polytechnique, ENSTA Paris, Institut Polytechnique de Paris, Palaiseau, France}

\author{I.A. Andriyash}
\affiliation{ LOA, CNRS, \'Ecole Polytechnique, ENSTA Paris, Institut Polytechnique de Paris, Palaiseau, France}

\author{A. Vernier}
\affiliation{ LOA, CNRS, \'Ecole Polytechnique, ENSTA Paris, Institut Polytechnique de Paris, Palaiseau, France}

\author{V. Tomkus}
\affiliation{Center for Physical Sciences and Technology, Savanoriu Ave. 231, LT-02300, Vilnius, Lithuania}

\author{V. Girdauskas}
\affiliation{Center for Physical Sciences and Technology, Savanoriu Ave. 231, LT-02300, Vilnius, Lithuania}
\affiliation{Vytautas Magnus University, K.Donelaicio St. 58. LT-44248, Kaunas, Lithuania}

\author{G. Raciukaitis}
\affiliation{Center for Physical Sciences and Technology, Savanoriu Ave. 231, LT-02300, Vilnius, Lithuania}

\author{M. $\mathrm{Mackevi\check{c}i\bar{u}t\dot{e}}$}
\affiliation{Center for Physical Sciences and Technology, Savanoriu Ave. 231, LT-02300, Vilnius, Lithuania}

\author{V. Stankevic}
\affiliation{Center for Physical Sciences and Technology, Savanoriu Ave. 231, LT-02300, Vilnius, Lithuania}

\author{A. Cavagna}
\affiliation{ LOA, CNRS, \'Ecole Polytechnique, ENSTA Paris, Institut Polytechnique de Paris, Palaiseau, France}

\author{J. Kaur}
\affiliation{ LOA, CNRS, \'Ecole Polytechnique, ENSTA Paris, Institut Polytechnique de Paris, Palaiseau, France}

\author{A. Kalouguine}
\affiliation{ LOA, CNRS, \'Ecole Polytechnique, ENSTA Paris, Institut Polytechnique de Paris, Palaiseau, France}

\author{R. Lopez-Martens}
\affiliation{ LOA, CNRS, \'Ecole Polytechnique, ENSTA Paris, Institut Polytechnique de Paris, Palaiseau, France}

\author{J. Faure}
\affiliation{ LOA, CNRS, \'Ecole Polytechnique, ENSTA Paris, Institut Polytechnique de Paris, Palaiseau, France}

\date{\today}

\begin{abstract}
We present significant advances in Laser Wakefield Acceleration (LWFA) operating at a 1\,kHz repetition rate, employing a sub-TW, few-femtosecond laser and a continuously flowing hydrogen gas target. We conducted the first comprehensive study assessing how the nature of the gas within the target influences accelerator performance. This work confirms and elucidates the superior performance of hydrogen in LWFA driven by few-cycle, low-energy laser pulses. Our system generates quasi-monoenergetic electron bunches with energies up to 10 MeV, bunch charges of 2 pC, and angular divergences of 15 mrad. Notably, our scheme relying on differential pumping enables continuous operation at kHz repetition rates, contrasting with previous systems that operated in burst mode to achieve similar beam properties. Particle-in-cell simulations explain hydrogen's superior performances: the ionization effects in nitrogen and helium distort the laser pulse, negatively impacting the accelerator performance. These effects are strongly mitigated in hydrogen plasma, thereby enhancing beam quality. This analysis represents a significant step forward in optimizing and understanding kHz LWFA. It underscores the critical role of hydrogen and the imperative need to develop hydrogen-compatible target systems capable of managing high repetition rates, as exemplified by our differential pumping system. These advances lay the groundwork for further developments in high-repetition-rate Laser Plasma Accelerator (LPA) technology.

\end{abstract}

\maketitle

\section{Introduction}

Laser wakefield acceleration (LWFA) is an alternative method for accelerating electrons to relativistic velocities \cite{tajima_laser_1979}. In LWFA, a high-intensity laser pulse generates an accelerating structure, the so-called wakefield, in a plasma. This allows to surpass the breakdown limit that exists in conventional radiofrequency (RF) cavities: the accelerating fields in the plasma typically reach amplitudes four orders of magnitude higher than those in RF cavities. Thus, acceleration occurs over very short distances, paving the way for compact accelerators.\\
Currently, Laser-Plasma Accelerators (LPAs) driven by 100\,TW to PW scale laser systems can accelerate electrons to energies ranging from $100\,$MeV \cite{faure_laserplasma_2004} to several GeV over a few centimeters \cite{kim_stable_2017, oubrerie_controlled_2022, miao_multi-gev_2022}. These high-energy LPAs demonstrated electron acceleration up to $8-10\,$GeV \cite{gonsalves_petawatt_2019, aniculaesei_acceleration_2023} using techniques such as laser-guiding or nano-particle injection. While pushing the energy frontier is important for high-energy physics, current research is also focusing on improving the accelerated beam quality and stability of these devices to enable their use in various applications. Indeed, electron beams from LPAs are being considered as drivers for femtosecond X-ray beams, either via betatron radiation \cite{rousse_production_2004}, Compton scattering \cite{ta_phuoc_all-optical_2012, chen_mev-energy_2013}, undulator radiation \cite{fuchs_laser-driven_2009}, or free electron laser radiation \cite{wang_free-electron_2021, khojoyan_transport_2016}. Such femtosecond X-ray beams could enable time-resolved (pump-probe) experiments based on e.g., X-ray diffraction or spectroscopy.\\
Research on stability improvement has been done for example in \cite{maier_decoding_2020}, where a continuous $24\,$h experiment was conducted, demonstrating high-quality statistics with $10^{5}$ consecutive electron beams generated at $1\,$Hz. Additionally, ongoing efforts are focused on developing high-repetition-rate LPAs ($100\,$Hz-kHz), which hold significant potential for improving stability, including the implementation of feedback loops within the accelerator system.\\
Until recently, kHz laser systems were limited to a few millijoules per pulse. However, recent progress in optical parametric amplification has enabled the development of higher energy systems \cite{budriunas_53_2017}. Ongoing research is pushing the boundaries further, with efforts to achieve Joule-level outputs in kHz systems using Ti:Sa technology. Additionally, new laser technologies, such as the coherent combination of fiber lasers, thin-disk lasers, and large-aperture Tm:YLF lasers, are being explored \cite{kiani_high_2023}. Still, the few-mJ systems remain compelling due to their compact design and turnkey operation, which make them particularly relevant for various societal applications.\\
In order to reach relativistic intensities for driving LWFA, the mJ laser pulses require sharp focusing and strong temporal compression, down to nearly a single optical cycle, i.e. a few femtoseconds \cite{guenot_relativistic_2017, gustas_high-charge_2018, salehi_mev_2017, faure_review_2018}. LPAs driven by few-cycle pulses can produce electron bunches in the MeV range with pC charge and potentially femtosecond durations.\\
Femtosecond electrons bunches in the $5-10\,$MeV range are relevant for various applications, including time-resolved electron diffraction \cite{he_electron_2013, faure_concept_2016}, high-dose rate radiobiology \cite{rigaud_exploring_2010, lundh_comparison_2012, cavallone_dosimetric_2021}, and the generation of ultrashort positron beams \cite{gahn_generation_2002}. For these applications, it is desirable for the accelerator to operate at kHz repetition rates. This poses technical challenges for the gas targets where the plasma is formed and acceleration occurs, as it is difficult to maintain a reasonable vacuum under these conditions.\\
In \cite{rovige_demonstration_2020}, Rovige \textit{et al.} demonstrated the continuous and stable operation of a kHz LPA for 5 hours using a free-flowing nitrogen jet. The continuous gas flow in the jet ensures a stable gas density profile even at a kHz repetition rate. Electrons had typical energies of a few MeV and stable beams were achieved thanks to shock injection in an asymmetrically shocked gas jet \cite{rovige_symmetric_2021}. Recently, Salehi \textit{et al.} managed to increase the electron energy up to 15 MeV by using a $\mathrm{H_2}$ plasma and kHz, few-cycle laser pulses \cite{salehi_laser-accelerated_2021}. However, they were limited to burst mode operation, as continuous operation led to a deterioration in accelerator performance, likely attributable to the rising background pressure inside the vacuum chamber. Even though the use of lighter gases is more demanding for the pumping system, this work shows that the use of $\mathrm{H_2}$ is beneficial in kHz LPAs driven by few-cycle pulses.\\
To gain more insight into the influence of the nature of the gas on the properties of the accelerated beams, we have performed a comprehensive study of the impact of the gas used to create the plasma, in the specific frame of few-cycle LWFA. In addition, we demonstrate an operational 1-kHz accelerator working with a continuous flow of $\mathrm{H_2}$, by maintaining a good vacuum in the chamber through a differential pumping scheme. This allowed us to systematically study the influence of three different gases on the accelerated electron beams: nitrogen $\mathrm{N_2}$, helium $\mathrm{He}$ and hydrogen $\mathrm{H_2}$.\\
The paper is organized as follows: in Sec. II, we recall the various known effects of the gas ionization on the laser pulse. Sec. III describes the differential pumping system and the experimental apparatus. In Sec. IV, we present the results of the gas comparison study, with additional emphasis on exploring the characteristics of electron beams accelerated with the $\mathrm{H_2}$ plasma. In Sec. V, the injection and acceleration process is investigated through particle-in-cell simulations, and we discuss our interpretations in Sec. VI. before delivering our conclusions in Sec. VII. 

\section{Ionization effects on the laser pulse}

In terms of energy balance, the ionization process can usually be neglected in LWFA, even for few-mJ pulses. In the particular case of our experiment, the absolute energy loss can be estimated as follows: the energy needed to ionize $\mathrm{N_2}$ up to the fifth level in a cylinder with a 10$\,\text{\textmu}$m radius and 200$\,\text{\textmu}$m length filled with gas with density $n_{N} = 2 \times 10^{19}\,\mathrm{cm}^{-3}$ is $E_{ion} = 0.05\,$mJ, which is only 1.8\% of the total energy of the laser pulse  $E_{laser} = 2.7\,$mJ.

\begin{figure}[h]
	\hspace{-0.5cm}
	\includegraphics[width=0.5\textwidth]{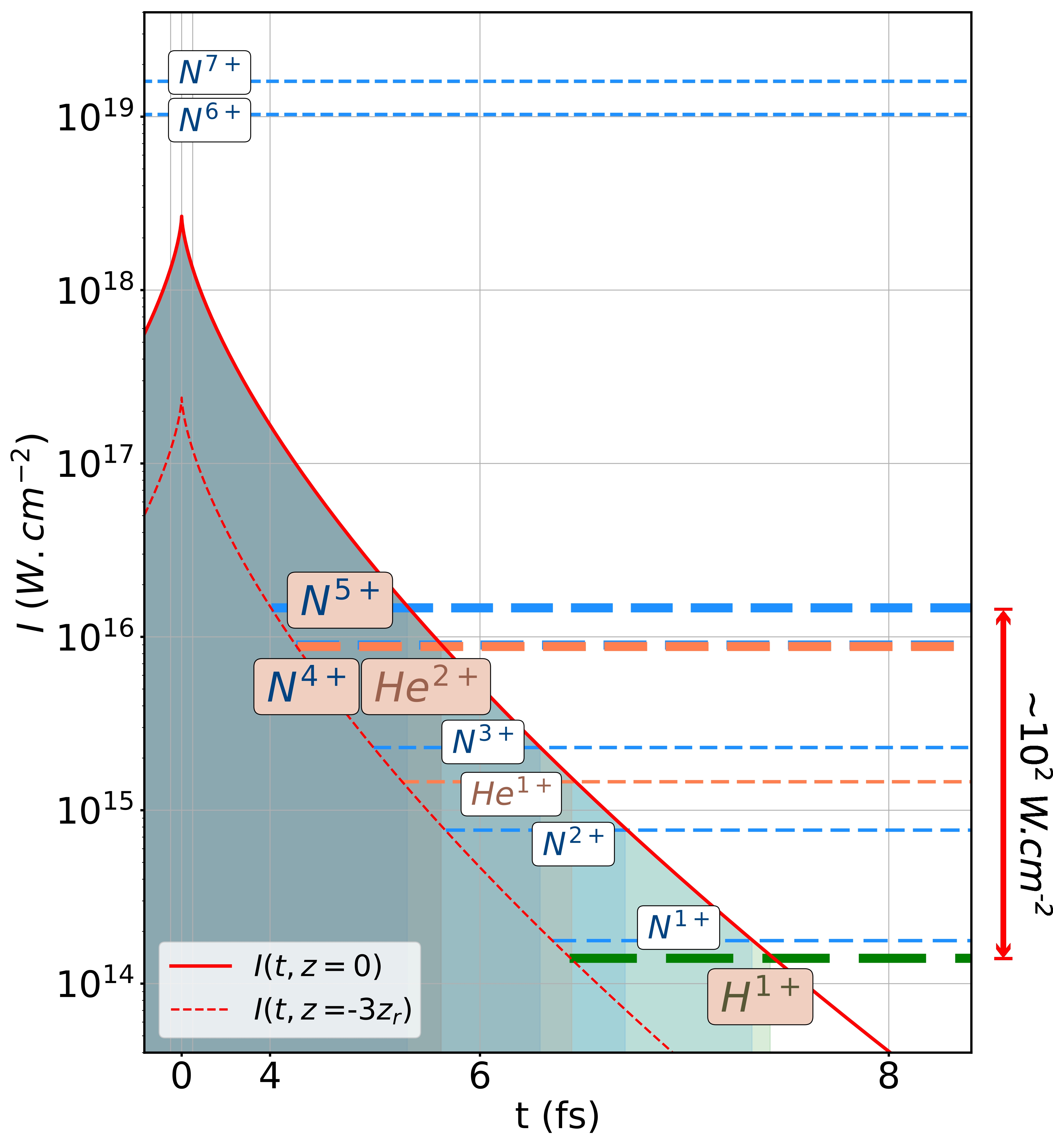}
	\caption{The laser intensity profile $I(t)$ at focus (solid red line) and away from focus at $z= - 3z_R$ (red dashed line). The horizontal lines correspond to the ionization threshold for the electrons of the different gases \cite{ammosov_tunnel_1986, augst_tunneling_1989}.}
	\label{fig:ion}
	\end{figure}

\noindent The main detrimental effects of ionization on the laser pulse come from the spatio-temporal distortions that are imprinted on the laser field. Indeed, the ionization rate varies in space and time. Figure \ref{fig:ion} shows the laser intensity temporal profile $I(t)$ for a Gaussian laser pulse with duration $\tau$ = $4\,$fs at FWHM, and peak intensity $I = 2 \times 10^{18}\,\mathrm{W\cdot cm}^{-2}$ at focus, as in our experiment. The dashed lines depict the barrier suppression intensity $I_{bs} = 4 \times 10^{9} E_i^4[eV] / Z^{*2}$ for each ionization level of the three gases \cite{ammosov_tunnel_1986, augst_tunneling_1989}. Here $E_i$ is the ionization energy and $Z^*$ the ionization level of the producted ion. The ionization occurs over very short distances (less than an optical cycle of the laser pulse): there is a front where ionization occurs. $\mathrm{H_2}$ is fully ionized when $I = 1.4 \times 10^{14}\,\mathrm{W\cdot cm}^{-2}$, between 7 and $8\,$fs before the peak intensity. Besides, the $2^\mathrm{nd}$ ionization level of $\mathrm{He}$ and the $4^\mathrm{th}$ and $5^\mathrm{th}$ levels of $\mathrm{N_2}$ are ionized when the laser reaches an intensity $I \simeq 10^{16}\,\mathrm{W\cdot cm}^{-2}$, two orders of magnitude higher than for $\mathrm{H_2}$. For a perfectly compressed laser pulse, this implies that the ionization front is at $5-6\,$fs, significantly closer to the intensity peak so that ionization of these levels will have a larger impact on the laser pulse distortion. This effect is even more pronounced when considering the laser before focus. The red dashed line in Figure \ref{fig:ion} shows the laser intensity profile at $z = -3z_R$ (which is only $\sim180\,\text{\textmu}$m in our experiment), where its intensity is decreased by a factor of 10. Consequently, the ionization front now appears $4\,$fs before the peak intensity. This suggests even stronger ionization effects that detrimentally impact the spatio-temporal profile of the pulse outside the focal region, as soon as the laser encounters significant gas density.

\noindent The first known effect of ionization is spectral/temporal distortion: an ultrafast variation of the refractive index $\eta$ results in  variations of the instantaneous frequency of the laser pulse \cite{yablonovitch_self-phase_1974, wilks_frequency_1988}:
\begin{equation}
    \Delta \omega = - \frac{\omega_0}{c} \int_0^L \frac{\partial \eta(\mathbf{r},t,z)}{\partial t}dz
\end{equation}
where $\omega_0$ is the central frequency, $\mathbf{r}$ the transverse coordinate and $z$ the propagation axis. The refractive index of the plasma is written $\eta = \left( 1 - \frac{\omega_p^2}{\omega_0^2}\right)^{1/2}$ where the plasma frequency is $\omega_p^2 = e^2n_e/m_e\epsilon_0$ with $n_e$ the electronic density of the plasma, $e$ and $m_e$ the electron charge and mass, respectively. This ionization-induced nonlinear phenomenon results in a frequency blueshift of the laser pulse \cite{wood_tight_1988, wood_measurement_1991}.

\noindent Secondly, in the transverse plane, the incoming laser pulse intensity decreases along the radial axis. Therefore, close to the propagation axis, the gas is uniformly ionized while it is less ionized on the edges of the beam. The electronic density is therefore non-uniform across the laser beam cross-section, which causes the plasma to act as a defocusing lens \cite{rae_ionization-induced_1993}. Then, the laser peak intensity at the focus is reduced, which negatively impacts the properties of the accelerated electron beam. 

\noindent In practice, both blue-shifting and ionization-induced defocusing combine to create spatio-temporal distortions of the laser pulse \cite{beaurepaire_limitations_2016}. Ionization occurs in a small region, smaller than an optical cycle, known as the ionization front. The laser pulse is typically eroded around the ionization front where these effects are strongest, causing the ionization front to shift closer to the  peak intensity of the pulse. While this can often be ignored for $>30\,$fs pulses with Joule energy, these effects become particularly critical when dealing with few-mJ few-cycle laser pulses: as the ionization front advances closer to the  peak intensity  within the laser pulse, ionization induced distortion becomes increasingly important. Consequently, $\mathrm{H_2}$ should greatly minimize these effects because it has a single ionization level two orders of magnitude lower compared to the other available gases, and should therefore be the preferred choice for LPA driven by few-mJ, few-cycle laser pulses.

\section{Experimental set-up}

\subsection{Differential pumping for using light gases in continuous operation}

\begin{figure}
	\centering
	\includegraphics[width=0.5\textwidth]{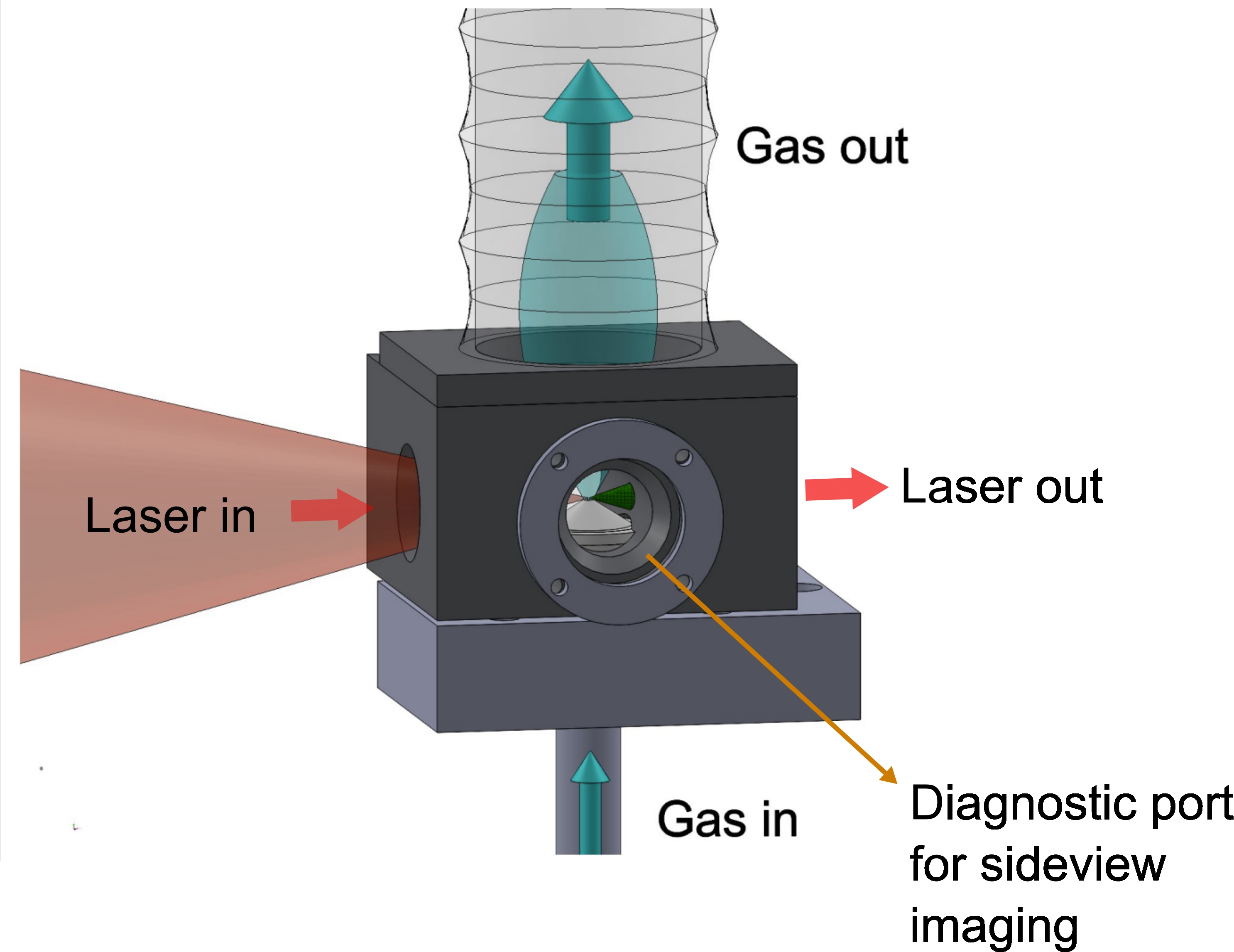}
	\caption{3D diagram of the differential pumping system. The jet is inside the chamber and two apertures let the laser pass through the chamber, above the gas jet.}
	\label{fig:pompdif}
\end{figure}

Even though $\mathrm{H_2}$ appears ideal to mitigate ionization effects, until now, kHz LWFA experiments with continuous operation have only been performed using $\mathrm{N_2}$ plasma \cite{guenot_relativistic_2017, gustas_high-charge_2018, faure_review_2018, rovige_demonstration_2020}. The use of $\mathrm{H_2}$ has been restricted to burst mode operation \cite{salehi_mev_2017, salehi_laser-accelerated_2021} because of pumping issues. Regarding the vacuum conditions, using $\mathrm{N_2}$ is more convenient because each molecule provides 10 electrons, while  atomic $\mathrm{He}$ or molecular $\mathrm{H_2}$ only provide two electrons. Therefore, in order to reach the same electron plasma density with $\mathrm{He}$ or $\mathrm{H_2}$, the backing pressure in the gas jet needs to be increased by a factor of about 5, which typically corresponds to $130-150\,$bar instead of $20-30\,$bar for $\mathrm{N_2}$ \cite{rovige_symmetric_2021}. When applying $20\,$bar of $\mathrm{N_2}$ in the gas jet, the gas flow rate in the experimental chamber is $Q = 12\,\mathrm{mbar \cdot l \cdot s^{-1}}$. Using $100\,$bar of $\mathrm{H_2}$ in the same gas jet, the leak rate becomes $Q = 224\,\mathrm{mbar \cdot l \cdot s^{-1}}$. Moreover, heavy gases are easier to pump than light gases: the compression ratio of the turbo-molecular pump goes as $K \propto \exp(\sqrt{M})$ with M the molar mass of the gas, and the pumping speed goes as $S \propto -\ln(K)$ \cite{gaede_diffusion_1915, becker_turbomolecular_1966}, so the lighter the gas, the slower the pumping speed of the turbomolecular pump. This has been the bottleneck for using continuous flows of $\mathrm{He}$ or $\mathrm{H_2}$ in LWFA.\\
\noindent To circumvent this issue, we have designed a differential pumping system enabling the use of light gases at high pressure with continuous flow. The differential pumping system is depicted in Figure \ref{fig:pompdif}. It consists of a small vacuum chamber positioned around the gas jet, with two holes to allow the propagation of the laser and the electrons through the chamber. A large dry screw pump with a maximum pumping speed of $650\,$m$^{3}$/h removes the gas directly from this small chamber just above the jet. This system makes it possible to keep the pressure in the main chamber below $10^{-3}\,$mbar while having a continuous flow of $\mathrm{He}$ or $\mathrm{H_2}$ from the nozzle with a backing pressure of up to $150\,$bar. 
\begin{figure*}[!t]

    \includegraphics[height=\textwidth, angle = -90]{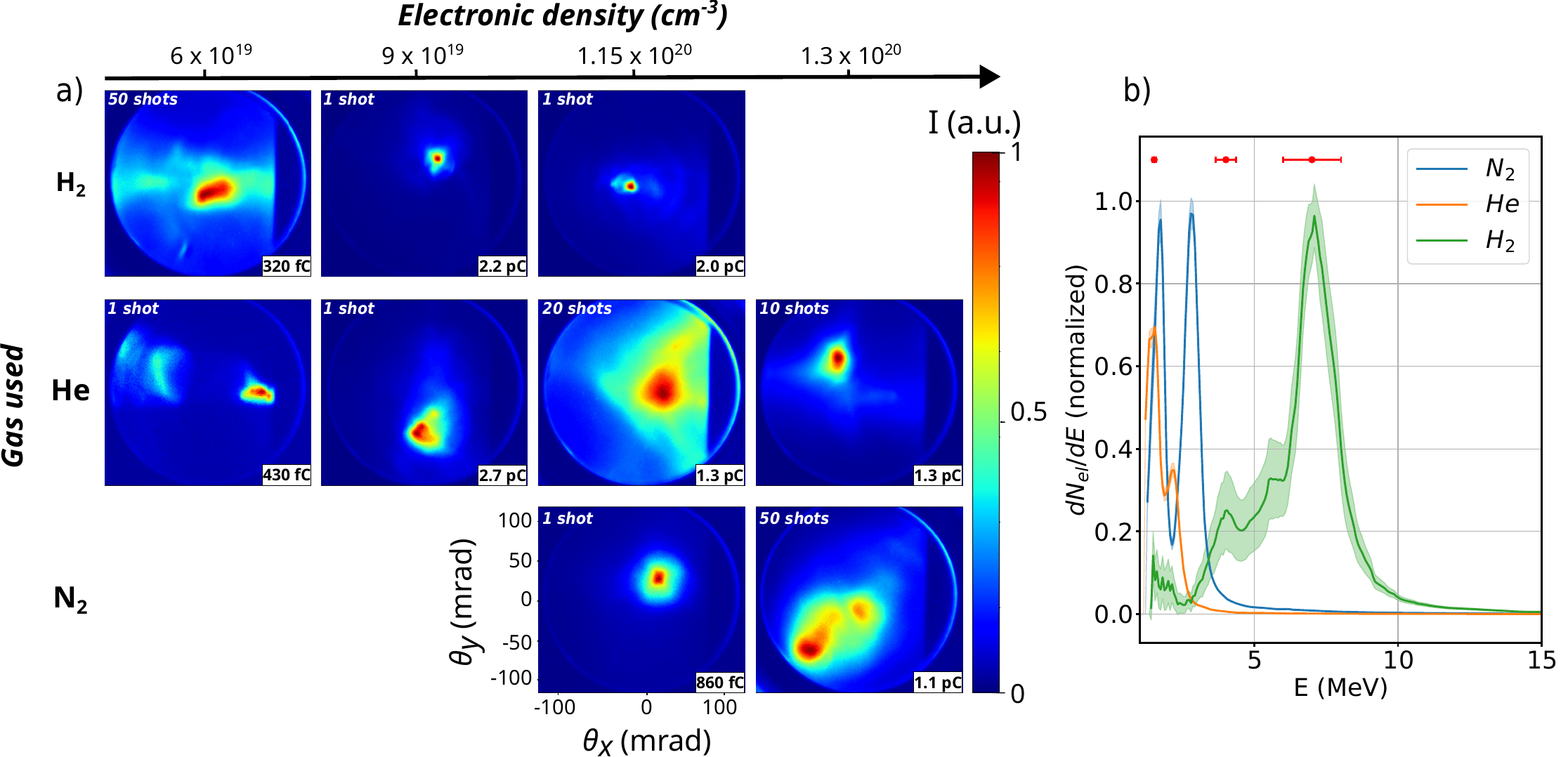} 
    \caption{Experimental results. (a) Variations in electron beam profile as the gas species changes for different values of the electron density. Each image presents an average of 1 to 50 shots, and the color scale is then normalized to its maximum level. (b) Spectra of the electron beam for $\mathrm{N_2}$, $\mathrm{He}$ and $\mathrm{H_2}$ with $n_e = 1.15 \times 10^{20}\,\mathrm{cm}^{-3} $. The thickness of the lines corresponds to the standard deviation from the mean value estimated over 10 acquisitions for $\mathrm{N_2}$ and $\mathrm{He}$, and 600 acquisitions for $\mathrm{H_2}$. The error bars (in red) correspond to the resolution of the spectrometer at $1.5$, $4$ and $7\,$MeV.}
    \label{fig:beam_prof}
\end{figure*}

\subsection{Laser wakefield accelerator}
\emph{Laser:} The laser wakefield accelerator is driven by the Salle Noire laser facility at the Laboratoire d'Optique Appliquée. It provides $10\,$mJ, $25\,$fs FWHM laser pulses at kilohertz repetition rate, with a central wavelength $\lambda_0 \sim 800\,$nm. The laser pulses are spectrally broadened in a hollow-core fiber filled with helium, which expands the spectrum from $650\,$nm to $900\,$nm FWHM through self-phase modulation. Subsequent compression is then achieved using chirped mirrors. The amount of dispersion in the pulse is controlled through a pair of motorized fused silica wedges, permitting to fine-tune pulse compression and measure the temporal intensity profile using the d-scan technique \cite{miranda_characterization_2012}. In the end, the laser delivers vertically polarized $4 \pm 0.2\,$fs pulses at FWHM with $2.7 \pm 0.08\,$mJ on-target energy at $1\,$kHz repetition rate \cite{bohle_compression_2014, ouille_relativistic-intensity_2020}. The pulses are focused by an off-axis parabola down to a $4.5 \pm 0.12\,\text{\textmu}$m spot at FWHM, reaching a peak intensity in vacuum $I = 1.8 \times 10^{18}\,\mathrm{W\cdot cm}^{-2}$. The Rayleigh length of the laser beam is $z_r$ = $62\,\text{\textmu}$m. The laser is focused at a distance of 150$\,\text{\textmu}$m from the exit of a continuously flowing supersonic-shocked gas jet with a 180$\,\text{\textmu}$m exit diameter \cite{tomkus_high-density_2018, marcinkevicius_femtosecond_2001}. The peak plasma density ranges within $n_e = (5 \pm 0.5) \times 10^{19}$ to $(1.5 \pm 0.15) \times 10^{20}\,\mathrm{cm}^{-3}$. The hydrodynamic shock within the supersonic gas jet creates a downward density gradient, which facilitates and stabilizes the injection of electrons into the wakefield \cite{rovige_demonstration_2020, rovige_symmetric_2021}.

\emph{Gas jet:} The gas jet used is a supersonic-shocked nozzle, as described in \cite{rovige_demonstration_2020, rovige_symmetric_2021}, with a throat diameter $60\,\text{\textmu}$m and an opening diameter $180\,\text{\textmu}$m. Density characterization is performed with $\mathrm{N_2}$ gas, as in \cite{rovige_symmetric_2021}: the gas column is illuminated from the side by white light and imaged on a quadriwave lateral shearing interferometer (SID4-HR, Phasics, \cite{primot_achromatic_1995, primot_extended_2000}). The electron density for a plasma made of $\mathrm{N_2}$ is then obtained assuming that each molecule is ionized up to the fifth level. In order to find the values for $\mathrm{H_2}$ and $\mathrm{He}$ we assume complete ionization (see Figure \ref{fig:ion}). In addition, we used a 1D isentropic expansion model for compressible gas flows to extrapolate the molecular density in $\mathrm{He}$ and $\mathrm{H_2}$. The molecular density in the flow $n_{mol}$, can be expressed according to the Mach number $M$ and its initial pressure and temperature, $P_0,\,T_0$ in the reservoir \cite{zucker_fundamentals_2002}: 
\begin{equation}
    n_{mol} = \frac{P_0}{k_BT_0}\left( 1 + \frac{\gamma - 1}{2}M^2\right) ^{-\frac{1}{\gamma - 1}}\\
\end{equation}
Here, the coefficient $\gamma=\frac{C_p}{C_v}$ is the specific heat ratio of the gas, and $k_B$ the Boltzmann constant. For our gas jet, where $M \gtrsim 3$, with $P_0$ = $40\,$bar, we obtain at the exit of the nozzle $n_{mol} = 3.18 \times 10^{19}\,\mathrm{cm}^{-3}$ for $\mathrm{N_2}$, $n_{mol} = 3.65 \times 10^{19}\,\mathrm{cm}^{-3}$ for $\mathrm{He}$ and $n_{mol} = 3.19 \times 10^{19}\,\mathrm{cm}^{-3}$ for $\mathrm{H_2}$. Thereafter and throughout this paper, the given electron density corresponds to the density at the plateau of the profile 150$\,\text{\textmu}$m above the gas jet. The shock has been characterized in previous work \cite{rovige_demonstration_2020,rovige_symmetric_2021} and is typically identified by a $20-25\%$ density bump compared to the plateau density.

\emph{Electron beam diagnostics:} The electron beam charge, spatial profile and pointing are measured using a calibrated YAG screen imaged onto a high dynamic range CCD camera (14-bit QImaging EXi Blue camera). The beam charge measurement has been cross-checked with an integrating charge transformer device \cite{unser_design_1989, meigo_measurement_2001}. Aluminum foils in front of the YAG screen block electrons with energies below $100\,$keV. The electron energy spectrum is measured onto the same YAG screen by inserting a motorized magnetic spectrometer into the beam. This spectrometer uses two different sets of permanent magnetic dipoles, with magnetic fields of $0.12\,$T and $0.4\,$T over a $20\,$mm length respectively, allowing different sensitivities on the electron spectrum.

\emph{Data acquisition}: Each image presented hereafter was obtained by averaging over 1 to 50 laser shots, depending on the signal level. For each measurement point (i.e.\,a given value of experimental parameters such as the density, the laser pulse duration, etc...), a series of 10 to 500 acquisitions was taken, depending on the required statistics. Statistics over electron beam parameters were then obtained by averaging over an acquisitions series, and the uncertainties represent the RMS deviation from the mean value.

\section{Results}
\subsection{Gas comparison}

We examined three different gases: nitrogen, helium and hydrogen. For each gas, a scan across plasma density $n_e$ was performed. For each measurement pair (gas, $n_e$), the beam charge and energy were optimized by adjusting the position of the gas jet with respect to the laser focal position, and by tuning the chirp of the laser pulse. During the optimization process, longitudinal position adjustments were made within a range of $\pm 0.6 z_r$, rendering them negligible.\\
The angular distributions of the electron beams are presented in Figure \ref{fig:beam_prof}a. The smallest beam divergence is reached in $\mathrm{H_2}$ with an electronic density $n_e = 1.15 \times 10^{20}\,\mathrm{cm}^{-3}$. It is nearly circular with a divergence of $\Delta \theta = 15.2$ ($\pm 7.0\,$mrad RMS) at FWHM in the laser polarization direction and $\Delta \theta = 13.0$ ($\pm 3.5\,$mrad RMS) FWHM in the perpendicular direction. For the same electronic density, the electron beam from $\mathrm{N_2}$  has a divergence $\Delta \theta = 49$ ($\pm 7.3\,$mrad RMS) FWHM (averaged in the two directions), which is 3.4 times bigger than in $\mathrm{H_2}$. In $\mathrm{He}$, for this particular density, the beam is even larger with a divergence $\Delta \theta = 120$ ($\pm 6\,$mrad RMS) FWHM. However, for higher density, the beam size in $\mathrm{He}$ goes down to $\Delta \theta = 42.5$ ($\pm 4.4\,$mrad RMS) FWHM while the beam size in $\mathrm{N_2}$ increases to $\Delta \theta = 76$ ($\pm 14.9\,$mrad RMS) FWHM. This density could not be reached in $\mathrm{H_2}$ as this would require backing pressures exceeding the limit of our pressure controller ($150\,$bar). The charge of the beam ranges from $1\,$pC to $3\,$pC for beams accelerated in $\mathrm{He}$ and $\mathrm{H_2}$ at densities higher than $9 \times 10^{19}\,\mathrm{cm}^{-3}$, but stays around $\sim 1\,$pC or less in $\mathrm{N_2}$.

\begin{figure}[hbtp]
     \centering
    \includegraphics[width = 0.5\textwidth]{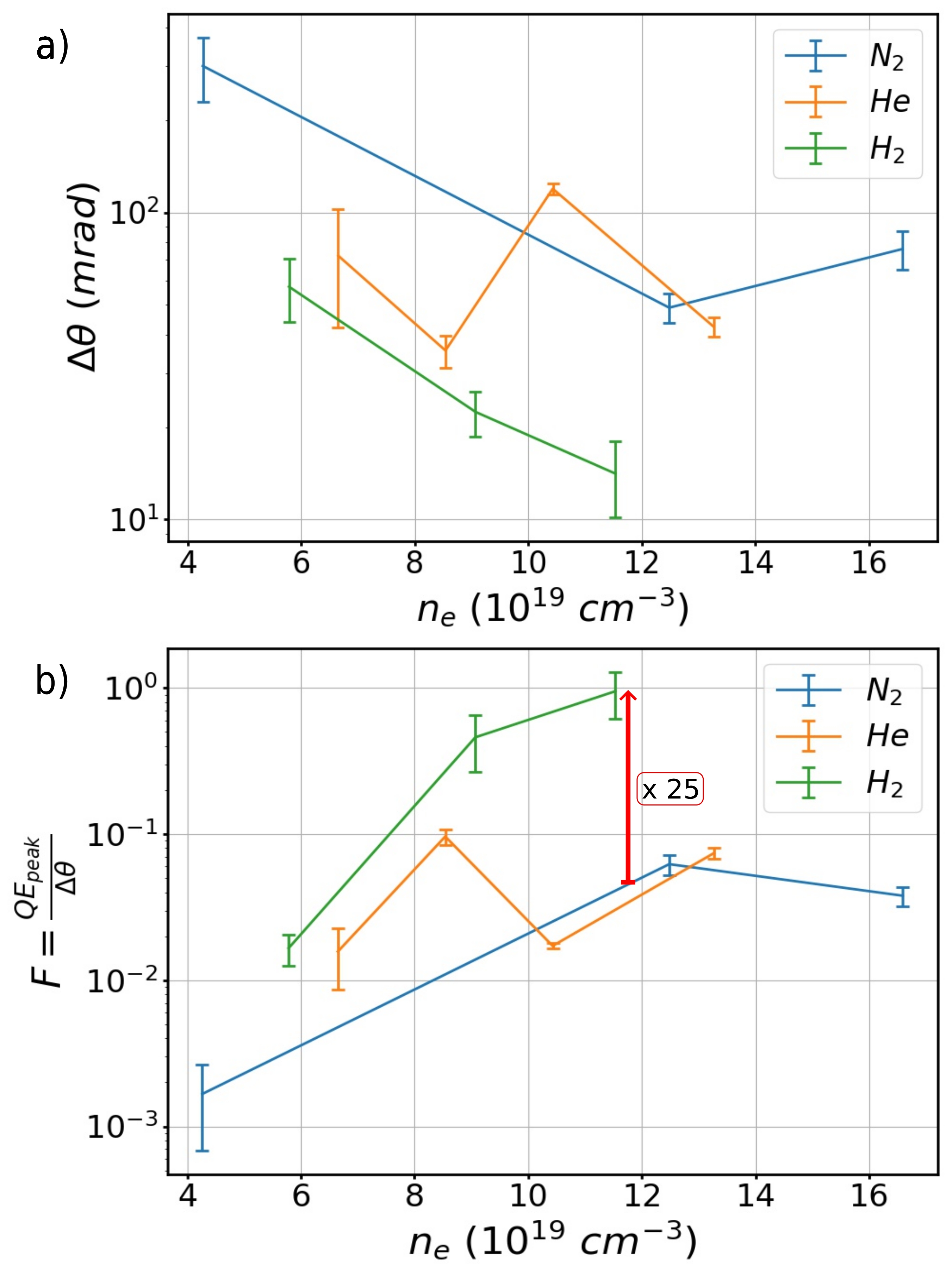}
    \caption{(a) FWHM beam divergence averaged on both the $x$ and $y$ axes, the smaller the value, the better the beam.  Each data point represents an average over at least 10 acquisitions. The vertical error bars indicate the standard deviation from the mean value. (b) Merit function $F(n_e,\ gas) = Q \times E_{peak} / \Delta \theta$, the higher the value, the better the beam. The vertical error bars are determined using the uncertainty propagation formula applied to the RMS values for each input parameter in $F$.}
    \label{fig:merit}
\end{figure}

\begin{figure*}[htbp]
     \centering
    \includegraphics[height = \textwidth, angle = -90]{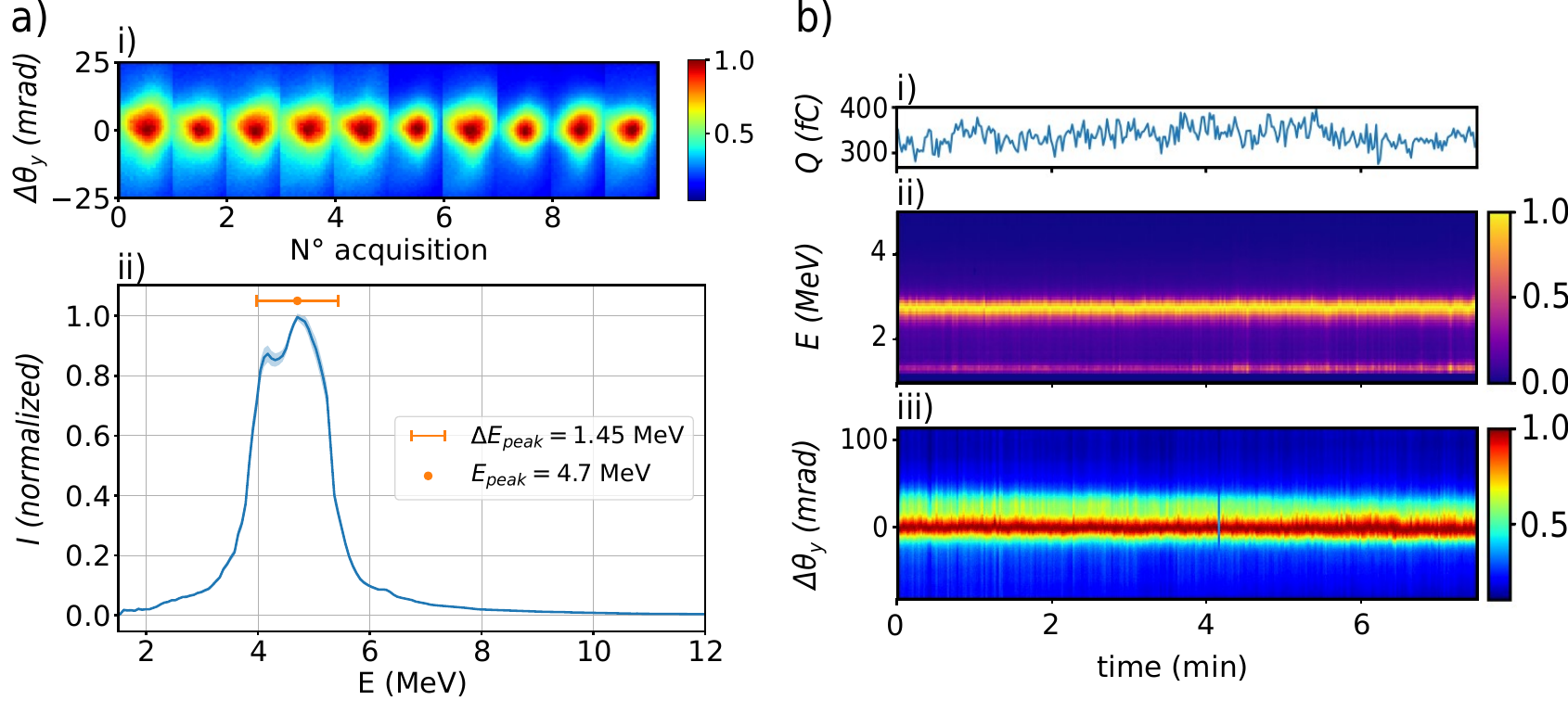}
    \caption{(a) Experimental results in a $\mathrm{H_2}$ plasma with $n_e = 6.6 \times 10^{19}\,\mathrm{cm}^{-3}$. (i) Electron beam profile. Each beam profile is obtained with 10 shots. (ii) Spectrum.  The spectrum is the result of averaging 100 shots. The thickness of the line corresponds to the RMS fluctuations of the spectrum. (b) Characteristics of the electron beam measured continuously over 300 shots (randomly sampled over 7 minutes) at density $n_e = 8 \times 10^{19}\,\mathrm{cm}^{-3}$. (i) Measured charge, each point is averaged over 50 shots. (ii) Measured electron spectra, each spectrum is averaged over 500 shots. (iii) Measured beam divergence, each profile is a slice of 3 pixels around the measured centroid of the beam averaged over 50 shots.}
    \label{fig:H2exp}
\end{figure*}

\noindent Figure \ref{fig:beam_prof}b, represents the spectrum (averaged over 10 acquisitions for $\mathrm{N_2}$ and $\mathrm{He}$, over 600 acquisitions for $\mathrm{H_2}$) of the electrons accelerated in the three gases with electronic density $n_e = 1.15 \times 10^{20}\,\mathrm{cm}^{-3}$. The thickness of the plot lines corresponds to the RMS fluctuations of the data. The red lines correspond to the resolution of the spectrometer. The electron beam accelerated with a $\mathrm{H_2}$ plasma clearly stands out with a peak energy $E_{peak} = 7\,$MeV and the tail of the distribution extending to $10-12\,$MeV. The width of the peak is limited by the resolution of the spectrometer. For the same density, electron beams from $\mathrm{N_2}$ show a peak at $2.5\,$MeV, and $\mathrm{He}$ shows two peaks at lower energy. We thus observe a threefold energy gain switching from a $\mathrm{N_2}$ or $\mathrm{He}$ plasma to a $\mathrm{H_2}$ plasma.\\
In order to compare more systematically the three gases, we implemented the following merit function: $F(n_e,\ gas) = \frac{Q \times E_{peak}}{\Delta \theta}$, $\Delta \theta$ being the mean of the FWHM of the beam along $x$ and $y$. This function promotes high-charge beams with a high energy peak and a low beam divergence. Figure \ref{fig:merit}a shows that the beam profile is smaller for electrons accelerated in $\mathrm{H_2}$ plasma. For low densities, the beam in $\mathrm{He}$ is smaller than in $\mathrm{N_2}$, but for higher densities their sizes becomes similar. Now looking at the merit function, once again $\mathrm{H_2}$ stands out with a merit function higher by more than an order of magnitude. Again, $\mathrm{He}$ yields better results than $\mathrm{N_2}$ for low densities.

\subsection{Exploring the stability of electron beams in $\mathrm{H_2}$ plasma}

Considering the clear advantage of $\mathrm{H_2}$ plasmas, we explored the properties of electron beams generated using a $\mathrm{H_2}$ plasma, focusing on obtaining high-quality and high-stability beams. Figure \ref{fig:H2exp}a shows the experimental results in a $\mathrm{H_2}$ plasma with $n_e = 6.6 \times 10^{19}\,\mathrm{cm}^{-3}$. The top part represents 10 acquisitions of the electron beam profile, each image being averaged over 10 shots. The electron beams have a charge per shot $Q = 970\,$fC ($\pm 30\,$fC RMS), which means that the relative RMS fluctuations of the charge shot to shot are as low as 3\%. The beam pointing has a RMS stability of $2\,$mrad, which means that the electron beam profile has excellent stability properties  considering its acceleration through LWFA. The lower part of Figure \ref{fig:H2exp}a shows the spectrum of the electrons averaged over 10 acquisition (each of 10 laser shots), and the thickness of the line corresponds to the RMS fluctuations of the spectrum. The spectrum is quasi-monoenergetic with a peak energy of $4.7\,$MeV ($\pm 0.1\,$MeV RMS), and with an energy spread of $1.45\,$MeV ($\pm 0.14\,$MeV RMS).\\
Figure \ref{fig:H2exp}b shows the stability study of the beam generated using a $\mathrm{H_2}$ plasma with a density $n_e = 8 \times 10^{19}\,\mathrm{cm}^{-3}$. We recorded 300 randomly sampled acquisitions over 7 minutes, while the accelerator operated at 1 kHz with no active stabilization. For each acquisition, the charge, spectrum, and spatial profile of the electron beam were recorded. The results show an averaged charge $Q = 340\,$fC with RMS relative fluctuations below 6\%, an averaged peak energy $E_{peak}$ = $2.7\,$MeV with a RMS stability as low as 0.7\%, and shot-to-shot beam pointing stability of only $2\,$mrad RMS.

\section{Simulations}

\begin{figure*}[htbp]
     \centering
    \includegraphics[width = \textwidth, angle = 0]{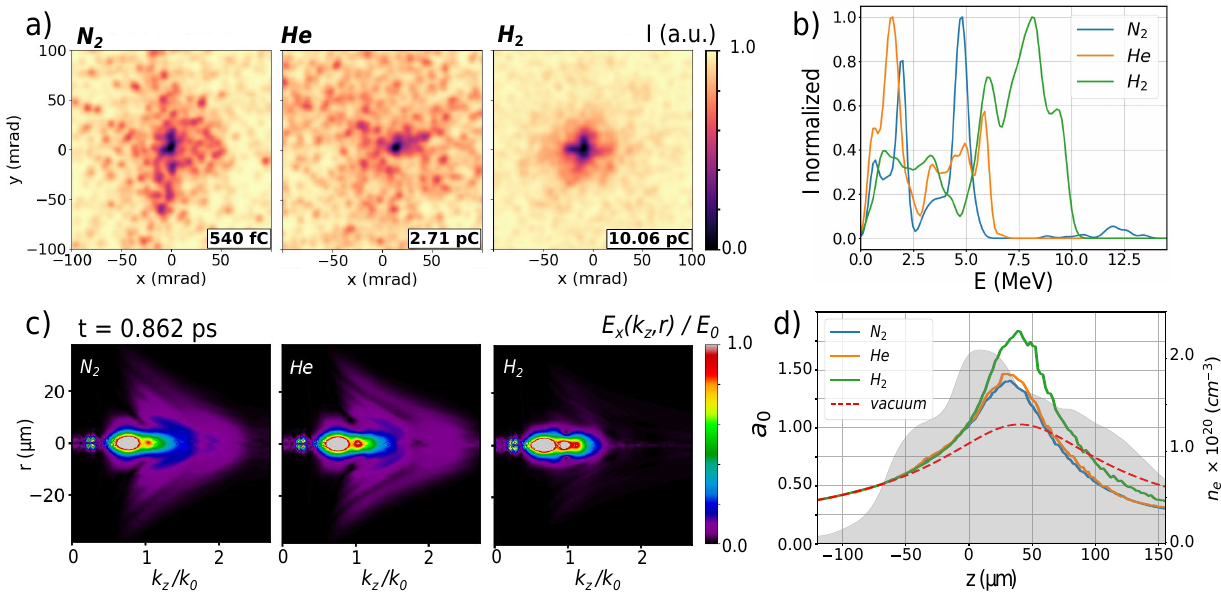}
    \caption{PIC simulations results. (a) Beam profile and (b) spectra of the simulated beams obtained in $\mathrm{N_2}$, $\mathrm{He}$ and $\mathrm{H_2}$ plasma with $n_e = 1.5 \times 10^{20}\,\mathrm{cm}^{-3}$, $E_{laser} =$ $2.7\,$mJ and $\tau = 4.2\,$fs. (c) Spatial spectrum of the laser pulse after propagation at $z = 75\,\text{\textmu}$m. The images are intentionally saturated to highlight the blue shift and defocusing. (d) Laser normalized amplitude $a_0$ along the propagation in the plasma $z$. The gray area represents the density profile of the gas used for the simulations. The density downramp is located at $z \simeq 30-40\,\text{\textmu}$m.}
    \label{fig:PICsim}
\end{figure*}

To understand the difference in performance of the accelerator when using different gases, we performed Particle-In-Cell (PIC) simulations for the three different gases while keeping the same laser parameters and final electron plasma density. The modeling is done with the spectral quasi-3D code FBPIC \cite{lehe_spectral_2016}, using a cylindrical simulation domain with the grid size $\Delta z = 20.6\,$nm, $\Delta r = 2 \Delta z$, three azimuthal Fourier modes, and 32 macro-particles of initially neutral gas per cell. The laser duration is $4.2\,$fs FWHM, with $2.7\,$mJ energy and a focal spot size of $4.5\,\text{\textmu}$m FWHM, corresponding to the experimental parameters. The gas target is initially neutral and consists of $\mathrm{N}$, $\mathrm{He}$, or $\mathrm{H}$ atoms, with number densities adjusted so that the final electron density is the same for all three cases. The density profile is taken from a previous measurement \cite{rovige_demonstration_2020} where a similar gas jet was used. The peak electron density is set to reach $n_e = 2 \times 10^{20}\,\mathrm{cm}^{-3}$ after ionization. Special care was taken to ensure that ionization was properly modeled as we expect the ionization process to be a key element to explain the experimental results: (i) all macro-particles are initially neutral and ionization relies on the ADK tunnel ionization model \citep{ammosov_tunnel_1986, augst_tunneling_1989}, (ii) the laser is initialized at $z=-180\,\text{\textmu}$m in the simulation so that the propagation of the laser pulse from the entrance of the gas jet is correctly accounted for (see \ref{fig:PICsim}d for the representation of the z-axis). Figure \ref{fig:PICsim} shows the result of the PIC simulation and displays the beam profiles (\ref{fig:PICsim}a) and the spectra (\ref{fig:PICsim}b) of the accelerated electron beams in the three gases. As expected, a higher charge and a better-defined spatial profile is reached in the case of a $\mathrm{H_2}$ plasma compared to the $\mathrm{He}$ or $\mathrm{N_2}$ case. Moreover, in $\mathrm{H_2}$, electrons achieve higher energies, with a peak at 8 MeV while $\mathrm{He}$ and $\mathrm{N_2}$ yield a 5 MeV energy peak. The simulations are in good agreement with the trends that are observed in experiments, with better results obtained with a $\mathrm{H_2}$ plasma in terms of charge, energy and divergence. 

\noindent These PIC simulations are now used to gain additional insight into the influence of ionization on the acceleration mechanism. The impact of propagation in a ionizing gas can be seen in Figure \ref{fig:PICsim}c. The electric field of the laser pulse is depicted as a function of the spatial frequency and radial position for the three gases at the time t$ = 0.862\,$ps, corresponding to propagation within the gas jet at $z = 75\,\text{\textmu}$m. At this point, the laser pulse has already crossed the density downramp of the jet. Compared to the $\mathrm{H_2}$ plasma, the laser pulse is significantly more distorted and shifted towards higher frequencies in $\mathrm{He}$ and $\mathrm{N_2}$ plasmas. This blue-shift can be directly attributed to gas ionization \cite{wood_tight_1988, wood_measurement_1991}. Note that ionization-induced defocusing is also visible in the figures for $\mathrm{He}$ and $\mathrm{N_2}$ as a transverse broadening of the laser pulse distribution, while it remains minimal for $\mathrm{H_2}$.

\begin{figure}[h!]
     \centering
    \includegraphics[width = 0.5\textwidth]{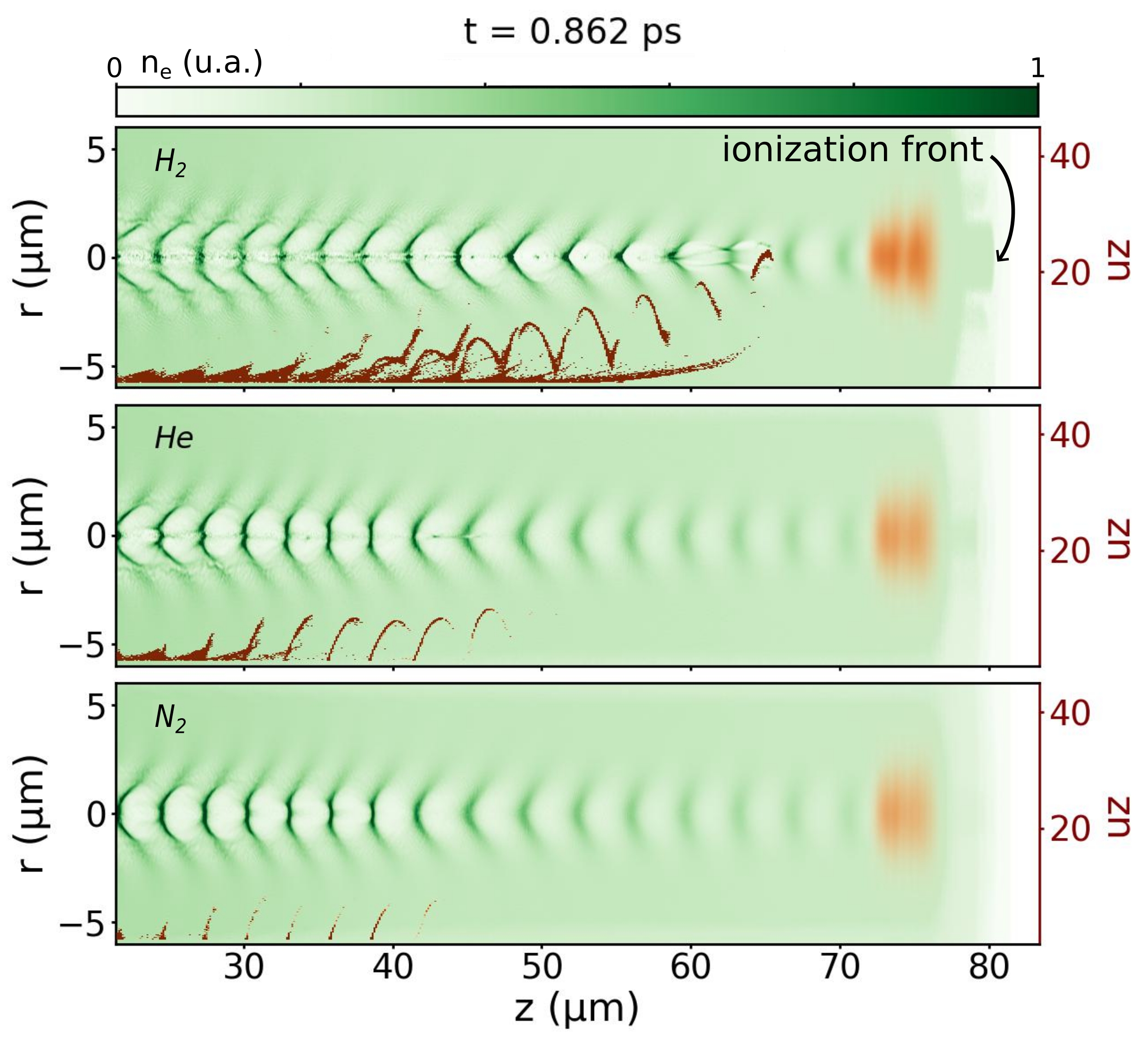}
    \caption{PIC simulation results at simulation time t = $0.862\,$ps. Images showing the laser pulse (orange), the plasma wakefield (electronic density in green) and the injected electrons longitudinal momentum $u_z=p_z/m_ec$ (brown) in longitudinal phase space. Each line correspond to the results in one gas, from top to bottom: $\mathrm{H_2}$, $\mathrm{He}$ and $\mathrm{N_2}$.}
    \label{fig:film}
\end{figure}

\noindent Figure \ref{fig:PICsim}d represents the normalized amplitude of the laser pulse, or normalized vector potential $a_0=eE/m_ec$, during propagation throughout the gas jet for the three gases under study, along with the theoretical $a_0$ for propagation in vacuum. In $\mathrm{H_2}$, $a_0$ reaches 1.8, while in $\mathrm{He}$ and $\mathrm{N_2}$, it reaches 1.5 and 1.4, respectively. Notably, $\mathrm{He}$ and $\mathrm{N_2}$ exhibit a very similar behavior. Consequently, self-focusing is more effective in $\mathrm{H_2}$ than in $\mathrm{He}$ or $\mathrm{N_2}$, as shown by the higher $a_0$ value. This suggests that the laser pulse experiences greater distortion in $\mathrm{He}$ and $\mathrm{N_2}$ compared to $\mathrm{H_2}$, which reduces the self-focusing effect in these gases.\\
Finally, Figure \ref{fig:film} illustrates the dynamics of electron injection for the different gases at time t = $0.862\,$ps. The wakefield within the electronic density (shown in green) is driven by the laser pulse here depicted in orange. Notably, the wakefield exhibits a larger amplitude in the $\mathrm{H_2}$ plasma compared to $\mathrm{He}$ or $\mathrm{N_2}$ plasma. Longitudinal momentum profiles of the injected electrons are plotted in brown as a function of longitudinal position. In the $\mathrm{H_2}$ plasma, electrons are injected around $10\,\text{\textmu}$m behind the laser pulse within the plasma density downramp. In contrast, in the $\mathrm{He}$ and $\mathrm{N_2}$ plasmas, electron injection occurs much later, $30$ to $50\,\text{\textmu}$m behind the laser pulse, and accelerated electrons reach lower longitudinal momenta than for the $\mathrm{H_2}$ case.

\section{Discussion}

The simulation results confirm that ionization effects induce strong distortions in the laser pulse, manifesting as a blueshift in the spectral domain and ionization-induced defocusing in the spatial domain. These effects degrade the wavefront of the laser pulse during propagation, diminishing its ability to self-focus. Notably, in a $\mathrm{H_2}$ plasma where ionization effects are relatively weak, the laser reaches higher intensity at focus, leading to the generation of a wake with a larger amplitude. \\
In a stronger wake, electrons are trapped more readily \cite{esarey_trapping_1995, faure_plasma_2016}. Consequently, in the $\mathrm{H_2}$ plasma, where ionization effects are relatively weak, the stronger wake facilitates rapid trapping of electrons.\\
In contrast, in the $\mathrm{He}$ and $\mathrm{N_2}$ plasmas, the wake amplitude is weaker because the laser intensity itself is weaker. Therefore, electrons are trapped further away from the laser pulse, at a location where the wakefield slows down. We recall the evolution of the wake phase velocity in a non-uniform plasma \cite{brantov_controlled_2008}: 

\begin{equation}
v_{\phi} = v_g \frac{1}{1 + (z - v_gt)\frac{1}{k_p}\frac{\partial k_p}{\partial z}}
\label{v_phase}
\end{equation}

\noindent With $k_p(z) = \sqrt{4 \pi r_e n_e(z)}$ the local plasma wave number, $r_e$ is the classical electron radius, and $(z - v_gt)$ is the the laser co-moving coordinate. When the distance from the laser pulse $(z_{las} - v_gt)$ increases, the plasma wave slows down, which facilitates the trapping of electrons.\\
The trapped electrons are injected in the wake with an initial $\gamma \simeq \gamma_\phi = \left( 1 - v_{\phi}^2/c^2\right)^{-1/2}$. The closer to the laser the electrons are injected, the higher $\gamma_\phi$. In addition, before entering the decelerating phase, the electrons travel the dephasing length $L_{deph}$ that can be approximated to $L_{deph} \simeq \gamma_\phi^2 \lambda_p$, where $\lambda_p$ is the plasma wavelength. Thus, the $L_{deph}$ grows quadratically with the initial momentum of the electrons $\gamma\sim\gamma_\phi$, which itself depend on the location of injection. In consequence, because injection occurs earlier in $\mathrm{H_2}$, the dephasing length is longer and electrons can reach higher energies.\\
In addition, the maximum energy gain is not only proportional to $L_{deph}$ but is also proportional to the amplitude of the electric field of the plasma wake:
\begin{equation}
	\Delta E \propto e E_{\textit{LWFA}} L_{deph}
\end{equation}
In a $\mathrm{H_2}$ plasma, the wake has a greater amplitude, so the accelerating field is higher, and the dephasing length is longer as well, as described above. This explains why the electrons accelerated in the $\mathrm{H_2}$ plasma reach higher energies in the experiments. On the contrary, in $\mathrm{He}$ and $\mathrm{N_2}$, the laser pulse is distorted because of ionization effects, thus self-focusing is weaker and the wake has a smaller amplitude. Electron injection occurs in buckets further away from the laser where the wakefield is slow enough to trap plasma electrons. Thus, accelerated electrons experience less intense accelerating fields for a shorter accelerating length, explaining the lower energies. Indeed, PIC simulations (see the electron longitudinal phase space in Figure \ref{fig:film}) confirm that the electrons injected far from the laser pulse dephase and undergo a decelerating field. In this case, the electrons might cross defocusing regions of the wake, which could also explain why they display a much larger divergence than the beams accelerated in the $\mathrm{H_2}$ plasma.

\section{Conclusion}

These results represent a clear experimental observation of the influence of the gas species on the electron beam quality and its spectrum in a kHz laser wakefield accelerator. Hydrogen clearly stands out from the other gases with higher energies (a 3-fold increase of the peak energy in $\mathrm{H_2}$ compared to $\mathrm{N_2}$), higher charge and smaller divergence (down to 13 mrad). Moreover, electrons accelerated in a $\mathrm{H_2}$ plasma display excellent beam pointing and spectrum stability, with a shot-to-shot beam pointing stability down to 2 mrad RMS, and a peak energy stable at 0.7\% RMS shot-to-shot. Electrons accelerated in $\mathrm{He}$ demonstrate results closer to those accelerated in nitrogen $\mathrm{N_2}$ than in hydrogen $\mathrm{H_2}$, primarily due to similar ionization levels occuring at $I = 10^{16}\,\mathrm{W\cdot cm}^2$. The implementation of the differential pumping system was a key enabler for this study, as it allowed us to conduct our experiments using light gases such as $\mathrm{H_2}$ to produce high-energy kHz electron beams in a continuous mode operation. PIC simulation results reproduce all of the experimental observation and confirm that ionization effects significantly distort the laser pulse in $\mathrm{He}$ and $\mathrm{N_2}$, which is found to be the cause of the loss of performance of the accelerator for these two gases.\\
Our results demonstrate that despite the challenges posed by its pumping system and safety considerations, $\mathrm{H_2}$ consistently outperforms other gases in generating high-quality electron beams due to its unique and low ionization level. These findings underscore the critical role of gas selection in laser-plasma acceleration, particularly when using few-cycle, few mJ laser pulses.

\section*{Acknowledgments}
This project has received funding from the European Union's Horizon 2020 research and innovation programme under grant agreement JRA PRISE no. 871124 Laserlab-Europe and IFAST under Grant Agreement No 101004730. This project was also funded by the Agence Nationale de la Recherche under Contract No. ANR-20-CE92-0043-01, and benefited from the support of Institut Pierre Lamoure via the chair “Accélération laser-plasma haute cadence”.\\
Financial support from the R\'{e}gion Ile-de-France (under Contract No. SESAME-2012-ATTOLITE) and the Extreme Light Infrastructure-Hungary Non-Profit Ltd (under Contract No. NLO3.6LOA) is gratefully acknowledged. We also acknowledge Laserlab-Europe, Grant No. H2020 EC-GA 654148 and the Lithuanian Research Council under Grant agreement No. S-MIP-17-79.

\section*{Data availability}
The data that support the findings of this study are available from the corresponding author upon reasonable request.

\bibliography{article_comp_gaz}

\end{document}